\documentclass[12pt]{article}
\usepackage{amsthm,amsmath,amsfonts,amssymb,ulem}
\usepackage{graphicx,psfrag,epsf,color,ulem,multirow,lineno,mathrsfs,algorithmic,algorithm2e,url}
\usepackage{enumerate}
\usepackage{natbib}
\usepackage[dvipsnames]{xcolor}
\usepackage{setspace}

\newcommand{\blind}{1}
\newtheorem{theorem}{Theorem}[section]
\newtheorem{definition}{Definition}[section]
\begin{document}

\def\spacingset#1{\renewcommand{\baselinestretch}%
{#1}\small\normalsize} \spacingset{1}

\if1\blind
{
  \title{\bf Regionalization of China's $\text{PM}_{2.5}$ through Robust Spatio-temporal Functional Clustering Method}
  \author{Tingyin Wang\hspace{.2cm}\\
    Department of Statistics and Finance, School of Management,  \\
    University of Science and Technology of China \\%Hefei, Anhui, 230026, China \\
   % christinawang666@mail.ustc.edu.cn\\
    and \\
    Xueqin Wang \thanks{
    Wang's research is partially supported by National Natural Science Foundation of China grants No.72171216, 12231017, 71921001, and 71991474, and the National Key R\&D Program of China No.2022YFA1003803.}\hspace{.2cm}\\
      Department of Statistics and Finance, School of Management,  \\
      University of Science and Technology of China\\ %Hefei, Anhui, 230026, China\\
     % wangxq20@ustc.edu.cn\\
      and\\
      Xiaobo Guo\\
      Department of Statistical Science, School of Mathematics,\\
        Southern China Research Center of Statistical Science, Sun Yat-Sen University\\
      %Guangzhou, Guangdong, 510275, China, \\
   %   guoxb3@mail.sysu.edu.cn\\
       and\\
      Heping Zhang \thanks{
    Zhang's research is supported in part by U.S. National Institutes of Health Grant Numbers
R01HG010171 and R01MH116527; National Science Foundation Grant Number DMS-2112711.}\hspace{.2cm}\\
     Department of Biostatistics, School of Public Health, Yale University\\}
     \date{}
  \maketitle
} \fi

\if0\blind
{
  \bigskip
  \bigskip
  \bigskip
  \begin{center}
    {\LARGE\bf Regionalization of China's $\text{PM}_{2.5}$ through Robust Spatio-temporal Functional Clustering Method}
\end{center}
  \medskip
} \fi

\newpage
\begin{abstract}
The patterns of particulate matter with diameters that are generally 2.5 micrometers and smaller ($\text{PM}_{2.5}$)
 are heterogeneous in China nationwide but can be homogeneous region-wide. To reduce the adverse effects from $\text{PM}_{2.5}$, policymakers need to develop location-specific regulations based on nationwide clustering analysis of $\text{PM}_{2.5}$ concentrations. However, such an analysis is challenging because the data have complex structures and are usually noisy. In this study, we propose a robust clustering framework using a novel concept of depth, which can handle both magnitude and shape outliers effectively and incorporate spatial information. We apply this framework to a $\text{PM}_{2.5}$ dataset and reveal ten regions in China that have distinct $\text{PM}_{2.5}$ patterns, and the homogeneity within each cluster is also confirmed. The clusters have clearly visible boundaries and enable policymakers to develop local emission control policies and establish regional collaborative systems to control air pollution in China.
\end{abstract}

\noindent%
\vfill

\newpage
\spacingset{1.45} % DON'T change the spacing!
\section{INTRODUCTION}
With the acceleration of industrialization, air pollution is becoming an issue of global concern. Among various types of air pollutants, fine particulate matters with aerodynamic diameters less than $2.5$ $\mu$m ($\text{PM}_{2.5}$) are most detrimental to health because they are small enough to travel deeply into the respiratory tract, penetrate the lung barrier, and affect the whole circulatory system. Several studies confirmed that chronic exposure to $\text{PM}_{2.5}$ increases the risk of developing cardiovascular diseases, respiratory diseases \citep{jia2017toxicity}, and may also leads to higher mortality. When annual $\text{PM}_{2.5}$ concentration reached 35 $\mu\text{g/m}^3$, one study reported that the mortality risk was increased by $15\%$ compared with 10 $\mu\text{g/m}^3$ \citep{tran2019threshold}. In  2015, $\text{PM}_{2.5}$ caused an estimated 4.2 million deaths globally, making it one of the five major health risk factors in that year \citep{pope2002lung}.

There is a disparity in the effect of poor air quality. Developing countries experiencing rapid urbanization and industrialization, such as China, suffer the most from air pollution. The air quality in China has been declining since the economic reforms began in 1978. In 2013, serious smog affected more than 800 million people  throughout China. Additionally, in January of that same year, the $\text{PM}_{2.5}$ concentration in various cities exceeded the World Health Organization's (WHO) recommended levels for good health by six times at approximately $70\%$ of the days \citep{world20202020}. In order to address the issue of high $\text{PM}_{2.5}$ concentration, the Chinese government devoted a great deal of human resources and financial support to collecting high-quality data. For instance, a vast national monitoring network has been gathering real-time high-quality measurements of air pollutant concentrations since 2013. The dataset has emerged as a crucial reference for the establishment of environmental policies and the development of emission control strategies. However, without  in-depth analysis,  observations from the data may not accurately portray the true state of air pollution.  One of the complicating factors is the heterogeneity across different regions. The $\text{PM}_{2.5}$ concentrations are heterogeneous nationwide but may be relatively homogeneous within a smaller region.  Thus,  it is important to develop a statistical method for $\text{PM}_{2.5}$ data that can unravel local emission patterns. This endeavor aligns with the ``coordinated inter-regional prevention and control efforts" initiated by China \citep{liang2021modeling}  and addresses a shortcoming in the existing policies such as the ones proposed by \citet{ChinaCouncilEnvironment2014},  which did not directly consider the local $\text{PM}_{2.5}$ levels.
 
Clustering analysis of the $\text{PM}_{2.5}$ data is an effective method to consider the localized characteristics of $\text{PM}_{2.5}$
\citep{wang2015spatial,liang2021modeling} by grouping the cities with similar patterns into the same clusters and identifying areas of interactions while classifying areas with distinct tendencies into separate clusters. Specifically, we model the observed $\text{PM}_{2.5}$ concentration data at each station as functional data by following \citet{ramsay2008functional}. High variability and the existence of outliers in the data present additional challenges in the analysis of the data. To ensure the reliability of our conclusions, we need to establish a robust clustering process. Some state-of-the-art robust clustering methods, such as DBSCAN and HDBSCAN, may take multiple stages to form stable clusters from noisy data \citep{campello2013density,harris2021elastic}. Other methods rely on the notion of depth, which determines the centrality and induces a center-outward ordering on sample functions. The depth-based procedures such as DBCA can handle outliers effectively \citep{jeong2016data}. Thus, we propose a novel functional depth, called angular depth, to deal with outliers and the infinite dimension of the functional data in developing our clustering framework. The angular depth takes into account both the shape and magnitude of the functional data in a computationally efficient manner.

In environmental science, clustering analysis is sometimes referred to as ``regionalization." It helps establish local environmental control strategies and forecast future environmental patterns. Traditional regionalization techniques include empirical orthogonal function (EOF) and its extended version, called rotated empirical orthogonal function (REOF) \citep{wang2015spatial}. These methods assume linearity among variables, determine the cluster boundaries subjectively, are sensitive to outliers, and lack the capability to accommodate the intercorrelation of spatial and temporal features. As a remedy, we also consider robust functional clustering methods such as distance-based trimmed K-means \citep{banerjee2012robust} and spatial-temporal clustering strategy such as mixture Gaussian model incorporated with Markov Random Field \citep{liang2021modeling,jiang2012clustering}. However, none of the listed methods adequately integrate the spatial and temporal features in a robust manner.

To address these shortcomings, in this paper, we establish a robust spatio-temporal clustering framework. Our clustering analysis of China's $\text{PM}_{2.5}$ dataset provides a regionalization across the country. Our identified regions manifest homogeneity of $\text{PM}_{2.5}$ levels, and we also find evidence of regional interactions. 

The remainder of this paper is organized as follows. In Section \ref{methodology}, we introduce an innovative depth notion called angular depth and propose a robust spatial clustering framework upon it. In Section \ref{simulation}, we present a simulation study to show the effectiveness and robustness of our method. In Section \ref{application}, we introduce a $\text{PM}_{2.5}$ dataset from China, discuss why a robust regionalization is needed, perform real data analysis, and summarize the main findings from clustering results. The article concludes with a discussion in Section \ref{discussion}.

\section{METHODOLOGY}\label{methodology}
In this section, we introduce a new concept of functional depth, called angular depth (AD), and propose a robust spatio-temporal clustering framework based on AD.

 We restrict our discussion to the \textcolor{blue}{functional} space $\mathscr{L}_2(\mathbb{R})$. A data point $x(v) \in \mathscr{L}_2(\mathbb{R})$ is a function defined on $\mathbb{R}$, which we simplify as $x$. The inner product of $x,y$ is defined as $\langle x,y\rangle=\int x(v)y(v)dv$, and the induced norm is $\|x\|=\sqrt{\int{x(v)^2}dv}$. Let $\Omega$ denote any underlying sample space. A random function $X$ on $\Omega$ is defined as a map: $X:\Omega\to\mathscr{L}_2(\mathbb{R})$, where for a fixed $\omega\in\Omega$, $X(\omega)$ corresponds to an element in $\mathscr{L}_2(\mathbb{R})$. A random function $X$ is said to follow a Gaussian measure (or denoted as Gaussian random variable) if for any $x\in \mathscr{L}_2(\mathbb{R})$, $\langle x, X\rangle$ has a Gaussian distribution. A Gaussian Hilbert space refers to a linear space of Gaussian random variables. Also, let $\mathbb{I}(\cdot)$ denote the indicator function and let $\operatorname{Acos}(x_1,x_2):=\arccos\left\{\frac{\langle x_1, x_2\rangle}{\|x_1\|\cdot\|x_2\|}\right\}$. 

\subsection{Angular Depth}

In this section, we present angular depth by using the projection technique and extending the work of \citet{kim2020robust} to functional space. $\mu(X),$ the measure of $X$, is assumed to be non-atomic throughout this paper.

\begin{definition}\label{def1}
Let $X$ be a random function in $\mathscr{L}^2(\mathbb{R})$ and $z$ be a fixed element in $\mathscr{L}^2(\mathbb{R})$. Define
$$D(z|X)=1-4\int\left[\mathbb{E}\mathbb{I}(\langle u,X\rangle\leq \langle u,z\rangle)-\frac{1}{2}\right]^2\mu(du),$$
where $\mu(\cdot)$ is a standard Gaussian measure defined on $\mathscr{L}^2(\mathbb{R})$.
\end{definition} 

Notice that  $\mathbb{E}\mathbb{I}(\langle u,X\rangle\leq \langle u,z\rangle)$ is the value of cumulative distribution function (CDF) of $\langle u,X\rangle$ at point $\langle u,z\rangle$. 
The closer a point's CDF is to $\frac{1}{2}$, the closer that point is to the median. We measure the outlyingness of $z$ on the projection direction $u$ with respect to $X$ with
\begin{equation}\label{equ1}
    \left[\mathbb{E}\mathbb{I}(\langle u,X\rangle\leq \langle u,z\rangle)-\frac{1}{2}\right]^2,
\end{equation}
By integrating $u$ of (\ref{equ1}) over the standard Gaussian measure $\mu$, we have a global measure of outlyingness of $z$ with respect to $X$. We then use the reverse of this value as a measure of centrality and formulate  $D(z|X)$ with an additive constant of 1 and a multiplier of 4 so that it is in the range of $[0,1]$. To avoid numerical integration, the following theorem offers a close form and is essential for computing $D(z|X).$

\begin{theorem}\label{thm1}
Suppose that $X_1$ and $X_2$ are two independent copies of $X$. Then,
$$D(z|X)=\frac{2}{\pi}\mathbb{E}\left[\text{Acos}(X_1-z,X_2-z)\right].$$
\end{theorem}

Next, we introduce robust mean function estimation called depth-based trimmed mean to estimate the main tendency of a given cluster so that it can be used to form relatively stable clusters with noisy data.

\begin{definition}
    For a given sample set $X_1,X_2,\cdots,X_n$, and $\alpha \in (0, 1),$ let the $\alpha$-quantile of angular depth be $D_{\alpha}(z|X)$. We define the angular depth-based $\alpha$-trimmed mean as:
    $$
    X_{\text{TM}} = \frac{\sum_{i=1}^n X_i\mathbb{I}(D(X_i|X)\geq D_{\alpha}(z|X))}{\sum_{i=1}^n \mathbb{I}(D(X_i|X)\geq D_{\alpha}(z|X))}.
    $$
\end{definition}
The $\alpha$-trimmed mean stands for the pointwise mean of the remaining samples after the removal of $100\alpha\%$ points with the lowest angular depth values. The intuition is that samples exhibiting low-depth values tend to be outliers within the sample set. The value of $X_{\text{TM}}$ depends on the trimming parameters $\alpha$, which we set at $\alpha=0.2$ throughout this paper as in \citet{lopez2006depth}. Next, we integrated this concept into our clustering framework.

\subsection{Robust Spatio-temporal Clustering}\label{method}
Now, we establish a robust spatio-temporal clustering (RSTC) framework for functional data using angular depth. Suppose that the random functions are defined in a time domain $\mathcal{T}$ and sampled from locations in a spatial domain $\mathcal{D}$. For clarity, assume that all sample functions are observed at the same time points $t_1,\cdots,t_l\in\mathcal{T}$. Let $Y(s_i,t_j)$ be the observation of the $i^{\text{th}}$ sample at time $t_j$ and location $s_i\in\mathcal{D}\subset \mathbb{R}^2$, where $i=1,2,3,\cdots,n$ and $j=1,2,3,\cdots,l$. Assume that there are a total of $K$ clusters, and the cluster membership $k\in\{1,2,3,\cdots,K\}$ of the $i^{\text{th}}$ curve is denoted as $z_i=k$, with $i=1,2,3,\cdots,n$.

RSTC contains an initialization step followed by an iterative step. The initialization step involves providing the cluster membership results for all samples based on our prior knowledge of spatial information, which in our application includes administrative divisions and geographical features. 

After the initialization step, we begin an iteration process by alternately updating each cluster's main tendency estimator and the cluster label of each sample. For the $m^\text{th}$ iteration, we estimate the 0.2-trimmed mean function for each cluster $k$, which is ${X^k_{\text{TM}}}^{(m)}$. And we assign a new cluster label $k_0$ to the $i^{\text{th}}$ sample (i.e., $z_i^{(m+1)}=k_0$) if ${X^{k_0}_{\text{TM}}}^{(m)}$ is the closest to that sample among the $K$ estimated trimmed means and the following spatial restriction is satisfied. The minimum spatial distance between sample $i$ and any sample within cluster $k$ must be smaller than a pre-specified threshold. This threshold must be chosen on \textcolor{blue}{a} case-by-case basis. It should be sufficiently small to promote spatial coherence while avoiding excessive reliance on the initial cluster memberships. After all samples have been examined and reassigned, we update the trimmed mean function ${X^k_{\text{TM}}}^{(m+1)}$ for $k=1,2,3,\cdots,k$ of each cluster based on $z_i^{(m+1)}$, where $i=1,2,3,\cdots,l$. 

Generally, the iteration process stops until no cluster memberships of samples are updated after the latest iteration. In special cases where the cluster memberships of samples undergo only minor changes after multiple iterations, we stop the iteration process if the results exhibit minimal changes after a new iteration. Specifically, we construct the stopping criterion through the concept of an adjusted Rand index (ARI) \citep{hubert1985comparing}, which measures the similarity between two clustering results. We stop the iteration process in the $k$-th iteration if the ARI between the clustering results produced by the $(k-1)$-th and $k$-th iterations \textcolor{blue}{does not} surpass a predetermined value. The clustering framework of RSTC is given in Algorithm \ref{RSTCalgo}.

\IncMargin{1em}
\IncMargin{1em}
\begin{algorithm}
\SetKwData{Left}{left}\SetKwData{This}{this}\SetKwData{Up}{up}
\SetKwFunction{Union}{Union}\SetKwFunction{FindCompress}{FindCompress}
\SetKwInOut{Input}{Input}\SetKwInOut{Output}{Output}
\Input{ $Y(s_i,t_{j})$, $i=1,2,\cdots,n$, $j=1,2,\cdots,l$, where $Y(s_i,t_{ij})$ is the discrete observation at time $t_{ij}\in\mathcal{T}$ on a random curve at location $s_i\in \mathcal{D}\in\mathbb{R}^2$.}
\Output{A new result of the cluster memberships of $n$ samples generated by the iteration process.}
\textbf{Initialize}: Let $m=0$ and assign cluster membership for each sample: $z_i^0=k$, $k=1,\cdots, K$ with $i=1,2,\cdots,n$ based on prior knowledge of spatial information. :\\
\BlankLine
\For{$m\geq 0$}{
  \For{$k\leftarrow 1$ \KwTo $K$}{
Estimate the trimmed mean function of cluster $k$, ${X^{k}_{\text{TM}}}^{(m)}$ based on the cluster membership $(z_1^{(m)}, z_2^{(m)}, \cdots ,z_n^{(m)}).$
  }
  \For{$i\leftarrow 1$ \KwTo $n$}{
Assign sample $i$ to cluster $k$ with the closest trimmed mean ${X^{k}_{\text{TM}}}^{(m)}$ to the random function of time of that sample, namely $z_i^{(m+1)}=k$, if certain distance condition is fulfilled.
  }
}
\If(){$(z_1^{(m+1)}, z_2^{(m+1)}, \cdots ,z_n^{(m+1)})=(z_1^{(m)}, z_2^{(m)}, \cdots ,z_n^{(m)})$}{
Stops.
}
\caption{The robust spatio-temporal clustering process.}\label{RSTCalgo}
\end{algorithm}\DecMargin{1em}

\section{SIMULATION}\label{simulation}
We conduct a simulation study to examine the performance of our proposed RSTC method on spatial-temporal clustering. We generate a synthetic dataset with samples that are spatially clustered. The curves of samples from different clusters have distinct mean functions but the same covariance structure. To assess robustness, \textcolor{blue}{outlier samples with randomly sampled spatial locations are added to the set, and their corresponding curve shapes deviate} from that of the pre-existing clusters. We evaluate the output with two indexes. The first one is the adjusted Rank index (ARI). The second is the standardized root mean square error (RMSE), defined as $\sqrt{\frac{\|\mu_k(t)-\hat{\mu}_k(t)\|^2}{\|\mu_k(t)\|^2}}$, which is used to evaluate if the estimated mean function reflects the true $\mu_k$. 

The simulation includes two main parts. In the first part, we closely follow the settings of homoscedastic case simulation studies carried out by \citet{liang2021modeling}. We repeat the simulation for $50$ times. For each repetition, we simulate the $n=156$ points with coordinates $s_1,s_2,\cdots,s_{156} \in \mathbb{R}^2$ as these are the points in the real dataset introduced in Section \ref{application} below that lie over the rectangular region ($107^{\circ}$E-125$^{\circ}$E, 28$^{\circ}$N-43$^{\circ}$N) in Northern China. We denote the longitude coordinate of the $i^{th} $ sample as $\text{lon}_i$ and the latitude coordinates as $\text{lat}_i$. The cluster memberships are simulated by generating a Markov random field using Gibbs sampling with $\nu=0.5$, $k=2$, and a multinomial distribution. For the $i^{th}$ sample, assuming that $k$ is the cluster membership, we generate the synthetic curve for each sample from the following functional model: 
$$Y_i(t)|(z_i=k)=\mu_k(t)+\sum_{q=1}^{Q}\gamma_{i,q}\phi_q(t)+\epsilon(t),$$ for $i\in\{1,\cdots,n\}$, $t\in\{\frac{1}{30},\frac{2}{30},\cdots,\frac{29}{30},1\}$, $Q=2$, and $k\in\{1,2\}$. 
We set the two cluster-dependent mean functions as $\mu_1(t)=1.8\cos(\pi t^2)$ and $\mu_2(t)=\cos(\pi t)$. $\phi_1(t)=\sqrt{2}\sin(2\pi t)$ and $\phi_2(t)=b_3(t)$, where $b_3(t)$ is the third basic function of the four-dimensional cubic spline basis $\mathbf{B}(t)$ defined by \citet{de1978practical}. $\mathbf{\gamma}_q's$, are generated through the isotropic exponential covariance structure $\text{cov}(\gamma_{i,q},\gamma_{i^\prime,q})=\sigma^2_{\gamma,q}\exp(-\|s_i-s^\prime_i\|/\phi)$ with $\phi=1$ and $(\sigma_{\gamma,1}^2,\sigma_{\gamma,2}^2)=(7,2)$. The error term $\epsilon_i$ is a white-noise process with variance $\sigma_\epsilon^2=0.4$. All the settings except the expression of $\mu_k(t)$ follow those of \citet{liang2021modeling}.

Now we compare RSTC with the following clustering methods: (a) k-means clustering; (b) a discriminative functional mixture model proposed by \citet{bouveyron2015discriminative} (funFEM); (c) a mixture model with spatially dependent random effects but independent cluster memberships (FMM); (d) the spatial-functional mixture model under a Markov random field (FMM-MRF).

We restrict the minimum distance between that sample and the potential cluster to be smaller than the $\alpha$ percentile ($10\%$, $20\%$, and $30\%$) of the between-sample distance. The results are denoted as RSTC1, RSTC2, and RSTC3, respectively. The results are summarized in Table \ref{simutb1}.

\renewcommand{\arraystretch}{1.5}
\begin{table}
\setlength{\abovecaptionskip}{1cm}
\caption{Means and standard deviations of the adjusted Rand index (the larger, the better) and the RMSE (the smaller, the better) using different clustering strategies, based on 50 simulation results. The standard deviations are given in parentheses.}
\label{simutb1}
\resizebox{16cm}{!}{
\begin{tabular}{ccccccccccccccccccc}
\\
\hline
 & k-means & funFEM & FMM & FMM-MRF &  RSTC1 & RSTC2 & RSTC3\\
\hline
 ARI & 0.15(0.09) & 0.100(0.10) & 0.565(0.45) & 0.50(0.47) &0.77(0.13) & 0.79(0.14) & 0.77(0.15) \\
%& & 0.085 & 0.099 & 0.452 & 0.468 & 0.132 & 0.137 & 0.150\\
\hline
RMSE & 0.96(0.10) & 0.96(0.11) & 0.88(0.11) & 0.86(0.10) & 0.78(0.07) & 0.79(0.07) & 0.79(0.07)\\
%& &  0.103 & 0.105 & 0.110 & 0.104 & 0.074 & 0.067 & 0.067\\
\hline
\end{tabular}
}
\end{table}

\begin{figure}[!ht]
      \begin{center}
      \includegraphics[width=0.8\textwidth]{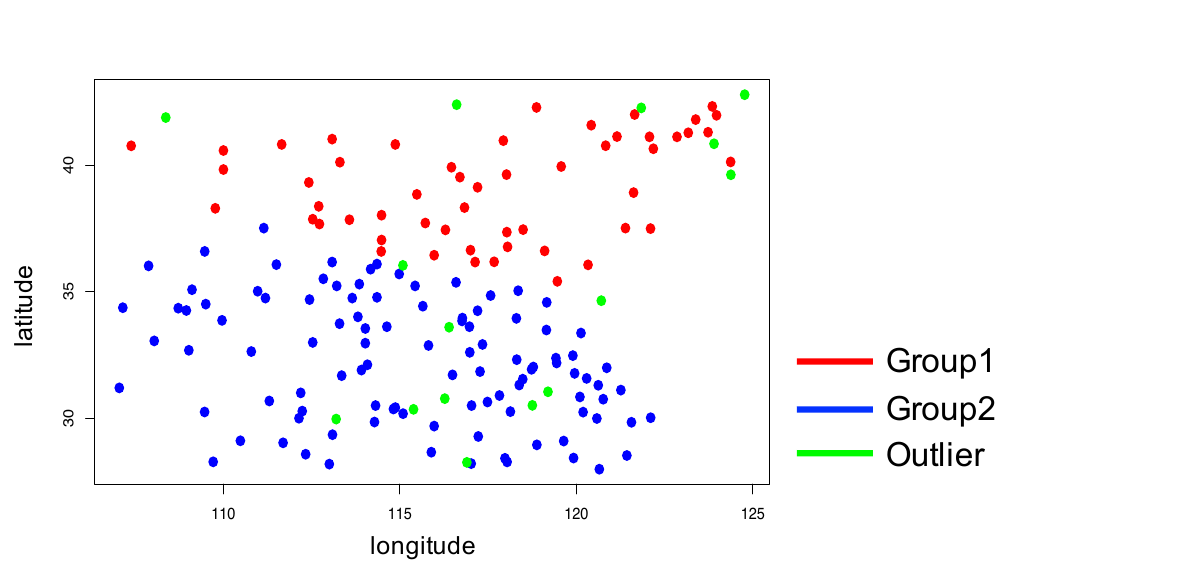} 
      \caption{The geographical locations of the $171$ samples in the rectangular region.}
      \label{simu3}
      \end{center}
   \end{figure}

\begin{figure}[!ht]
      \begin{center}
      \includegraphics[width=1.1\textwidth]{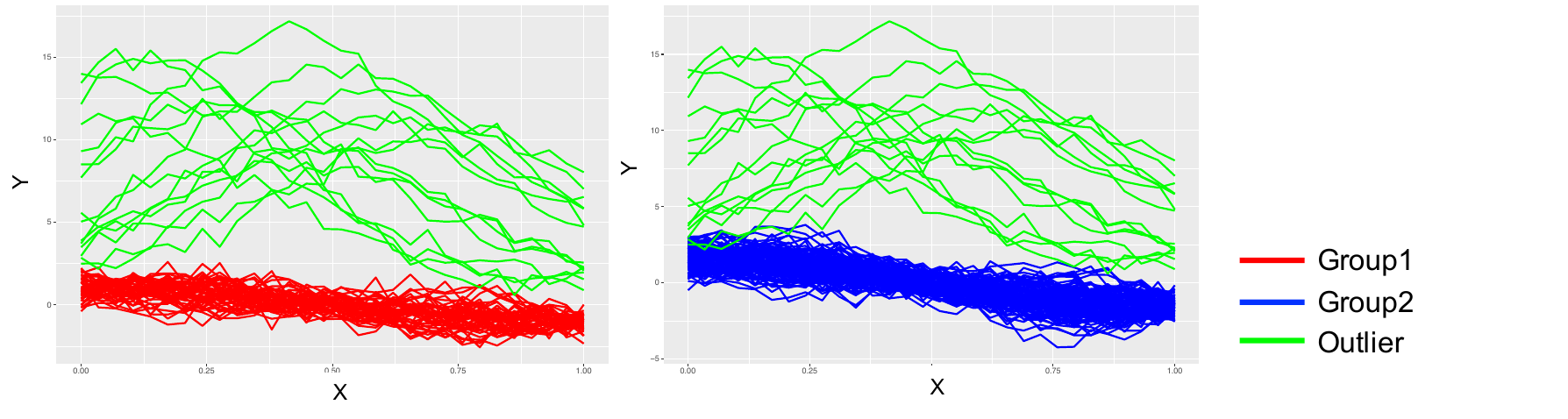} 
      \caption{Left: The curve shapes of the samples that belong to the first cluster and the 15 green outlier curves. Right: The curve shapes of the samples that belong the second cluster and the 15 green outlier curves.}
      \label{simu2}
      \end{center}
   \end{figure}

\begin{figure}[!ht]
      \begin{center}
      \includegraphics[width=0.6\textwidth]{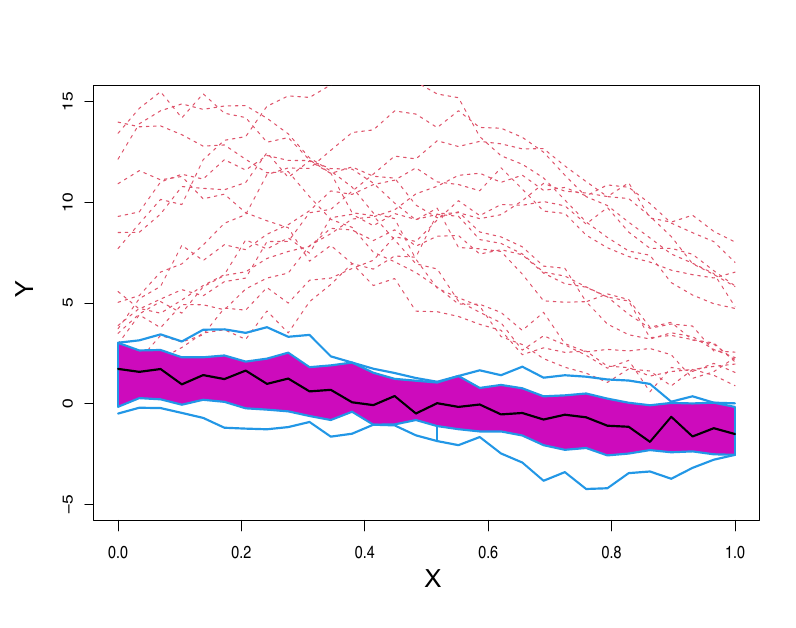} 
      \caption{ A Functional boxplot of all curves. The dashed red curves are identified as outliers by the boxplot.}
      \label{simu}
      \end{center}
   \end{figure}

In the second part of the simulation, the cluster membership is given by $z_i=1+\mathbb{I}(a_1\cos(2 \text{lon}_i) + a_2 \text{lon}_i\times \text{lat}_i > \tilde{a}_{\beta})$, with $a_1\sim U(-1,1)$, $a_2\sim U(-0.5,0.5)$, and $\tilde{a}_{\beta}$ is the $\beta-$quantile of $\{a_1\cos(2 \text{lon}_i) + a_2 \text{lon}_i\times \text{lat}_i,i=1,\cdots,n\}$, with $\beta\sim U(0.3,0.7)$. We \textcolor{blue}{first} generate $156$ \textcolor{blue}{curve samples, whose curve shapes and locations follow} the first part of the simulation. \textcolor{blue}{We then} add $15$ \textcolor{blue}{outlier samples} to the original \textcolor{blue}{sample} set. \textcolor{blue}{In constrast to the original $156$ samples, the locations of the outlier samples are randomly sampled from the rectangular region as illustrated in Figure \ref{simu3}, and their curve shapes are highly distinct from both each other and the original samples as illustrated in Figure \ref{simu2}.} 
\textcolor{red}{(What does ``spatial distribution" mean? I tried to  use ``spatial distribution" to refer to the locations of the samples, but I change my expression in the new version for clarity.) (Are the dots just the scatter plots? Yes, the scatter plots are used to show geographical locations of the generated samples in the rectangular.) (What is ``the sample" here? ``The sample" is a very unclear expression, and I change them into the ``outlier sample" and the ``original sample". Each sample refers to a curve that has a specific gepgraphical location.)}
\textcolor{blue}{The curves of the outlier samples are generated} as follows: 
$$Y_i(t)=3c\exp\{\cos\pi(t-c_1)\} + \sum_{q=1}^{Q}\gamma_{i,q}\phi_q(t)+\epsilon(t),$$
where $c \sim U(1,2)$  and $c_1 \sim U(0,1)$. \textcolor{blue}{To show their outlyingness,} we construct a functional boxplot for all curves following \citet{sun2011functional}, and it identifies the $15$ curves as outliers \textcolor{blue}{as displayed in Figure \ref{simu}}.  

The minimum distance restriction of RSTC is set as $\alpha$ percentile ($10\%$, $5\%$, and $2\%$) of the between-sample distance, and the results are denoted as RSTC1, RSTC2, and RSTC3.

 Table \ref{simutb1} shows RSTC method can perform better than other methods (including spatial-temporal methods such as FMM and FMM-MRF) when spatial correlation is strong and the distinction of time series between two clusters is weak. The \textcolor{blue}{advantage} of our method is that \textcolor{blue}{it} does not require pre-specified model assumptions\textcolor{blue}{, and performs }equally well as model-based strategies \textcolor{blue}{with correct pre-specified model assumptions.}

Table \ref{simutb3} shows that RSTC is the only method that can perform robustly in the presence of outliers. Unlike the other methods, RSTC is not prune to outliers and successfully identifies the two underlying clusters regardless of the existence of outliers. This is because RSTC recognizes the shapes of the 15 generated outlier curves as being spatial dispersed and heterogeneous as illustrated in Figures  \ref{simu3} and \ref{simu2}.

\begin{table}
\setlength{\abovecaptionskip}{1cm}
\caption{Means and standard deviations of the adjusted Rand index (the larger, the better) and the RMSE (the smaller, the better) using different clustering strategies, based on 50 simulation results. The standard deviations are given in parentheses.}
\label{simutb3}
\resizebox{16cm}{!}{
\begin{tabular}{ccccccccccccc}
\\
\hline
& k-means & funFEM & FMM & FMM-MRF &  RSTC1 & RSTC2 & RSTC3\\
\hline
ARI & 0.00(0.00) & 0.05(0.12) & 0.06(0.24) & 0.02(0.14) & 0.98(0.04) & 0.98(0.04) & 0.94(0.07) \\
 %&  & 0.000 & 0.008 & 0.000 & 0.000& 0.203 & 0.194 & 0.155 \\
\hline 
RMSE & 2.72( 0.75) & 2.72(0.75) & 2.68(0.82) & 2.72(0.75) & 0.71(0.14) & 0.67(0.12) & 0.65(0.12) \\
%  &  & 0.096 & 0.096 & 0.096 & 0.096 & 0.530 & 0.001 & 0.002 \\
  \hline
\end{tabular}
}
\end{table}

\section{APPLICATION TO REAL DATA SET}\label{application}
\subsection{Data Description}
The dataset contains China's city-level daily $\text{PM}_{2.5}$ concentration, which was processed and published by \citet{liang2021modeling}. The data were collected by China's Ministry of Ecology and Environment to assess air quality data through an extensive monitoring network from 2013. This network expanded to over 1,500 monitoring stations in 338 cities in 2015-2016. Real-time measurements of pollutants are continuously recorded and then transmitted to China National Environmental Monitoring Center (CNEMC). The data  were collected through either tapered element oscillating microbalance or Beta ray attenuation \citep{wang2015spatial},  and all devices met the CNEMC standards for continuous automated air pollutant measurement methods.

We take the average hourly $\text{PM}_{2.5}$ concentration data from all monitoring stations in each city to obtain the daily city-level $\text{PM}_{2.5}$ data. Each of the $338$ cities has $731$ measurements from January 1, 2015, to December 31, 2016, but $31$ out of them are removed for low quality. For the remaining $307$ cities, missing data $(<0.6\%)$ are imputed using linear interpolation following the suggestion of \citet{liang2021modeling}. The topographic information, namely the longitude and latitude coordinates for each city, is available. Pairwise distances between two locations are calculated using the geodesic distance, defined by \citet{kimmel1998computing} as the shortest distance between two points on the surface of a sphere.

\begin{figure}[!ht]
      \begin{center}
      \includegraphics[width=1.0\textwidth]{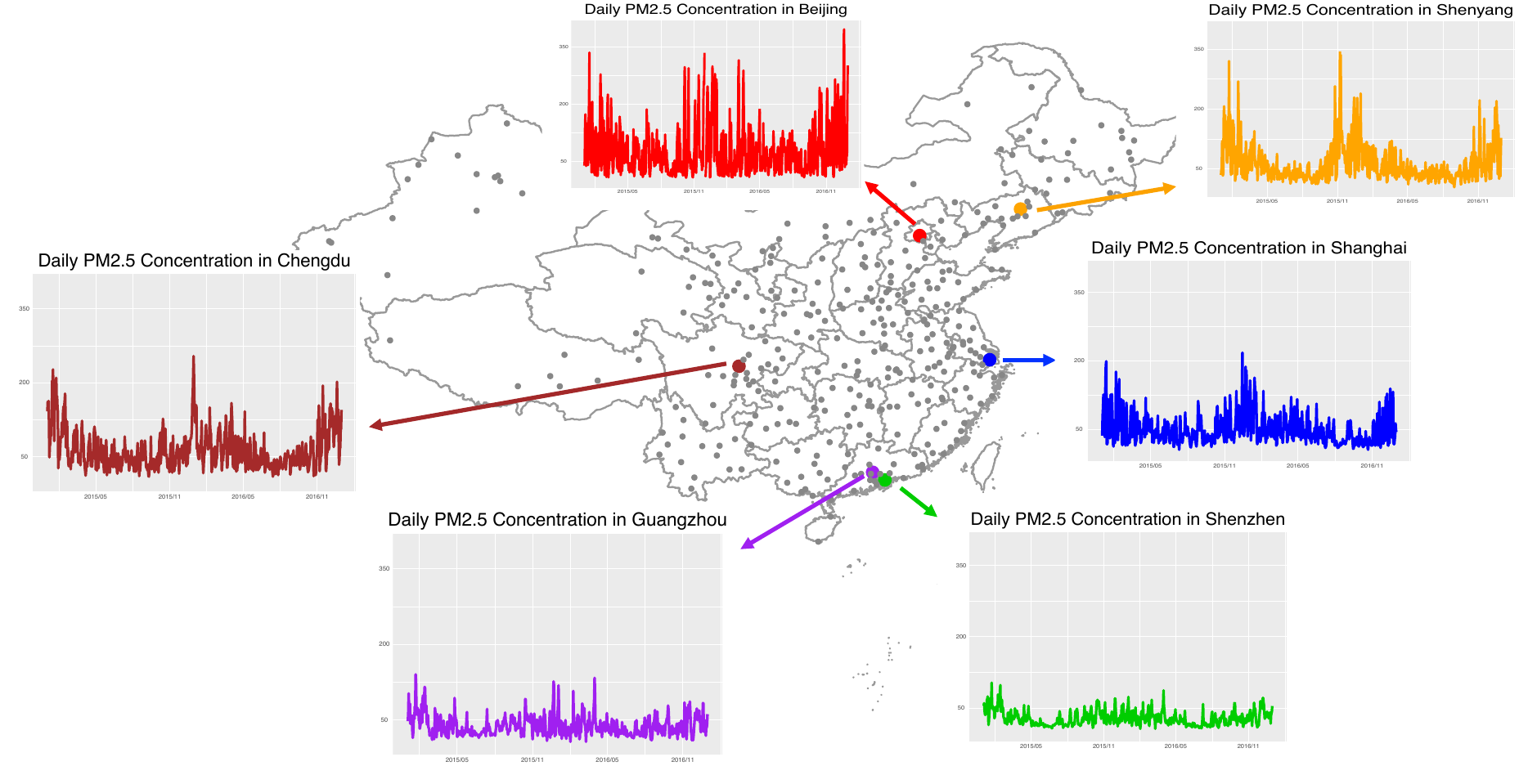} 
      \caption{Locations of all cities and the time series of $\text{PM}_{2.5}$ levels of six representative cities. 338 cities in China with monitoring stations are marked by dots. The times series plots of $\text{PM}_{2.5}$ daily concentrations($\mu\text{g/m}^3$) of Beijing, Shenyang, Shanghai, Chengdu, Guangzhou, and Shenzhen from January 2015 to December 2016 are also displayed.}
      \label{display}
      \end{center}
   \end{figure}

We display the locations of all 338 cities in Figure \ref{display}. For demonstration purposes, we highlight six main cities and display their average daily observations from January 2015 to December 2016. Among the six megacities, the $\text{PM}_{2.5}$ concentration levels of Guangzhou and Shenzhen are highly correlated throughout the year, mainly due to their geographical closeness. Yet, other cities exhibit apparent variations despite being almost the same latitude. For example, Chengdu has higher $\text{PM}_{2.5}$ levels than Shanghai, especially during the winter. One possible explanation is that Chengdu is located in a basin surrounded by mountains, which prevents the pollutants from scattering and allows them to accumulate more easily. In contrast, Shanghai is located on the coast, and the prevailing winds can disperse pollutants more efficiently and lower $\text{PM}_{2.5}$ concentration. Likewise, Beijing has much higher $\text{PM}_{2.5}$ levels than Shenyang. Higher population density and greater industrial activities in Beijing together may contribute to the phenomenon. These observations reveal that $\text{PM}_{2.5}$ patterns are heterogeneous across the country but rather homogeneous within a smaller area.

\subsection{Application of RSTC to Data}
\begin{figure}[!ht]
      \begin{center}
      \includegraphics[width=0.6\textwidth]{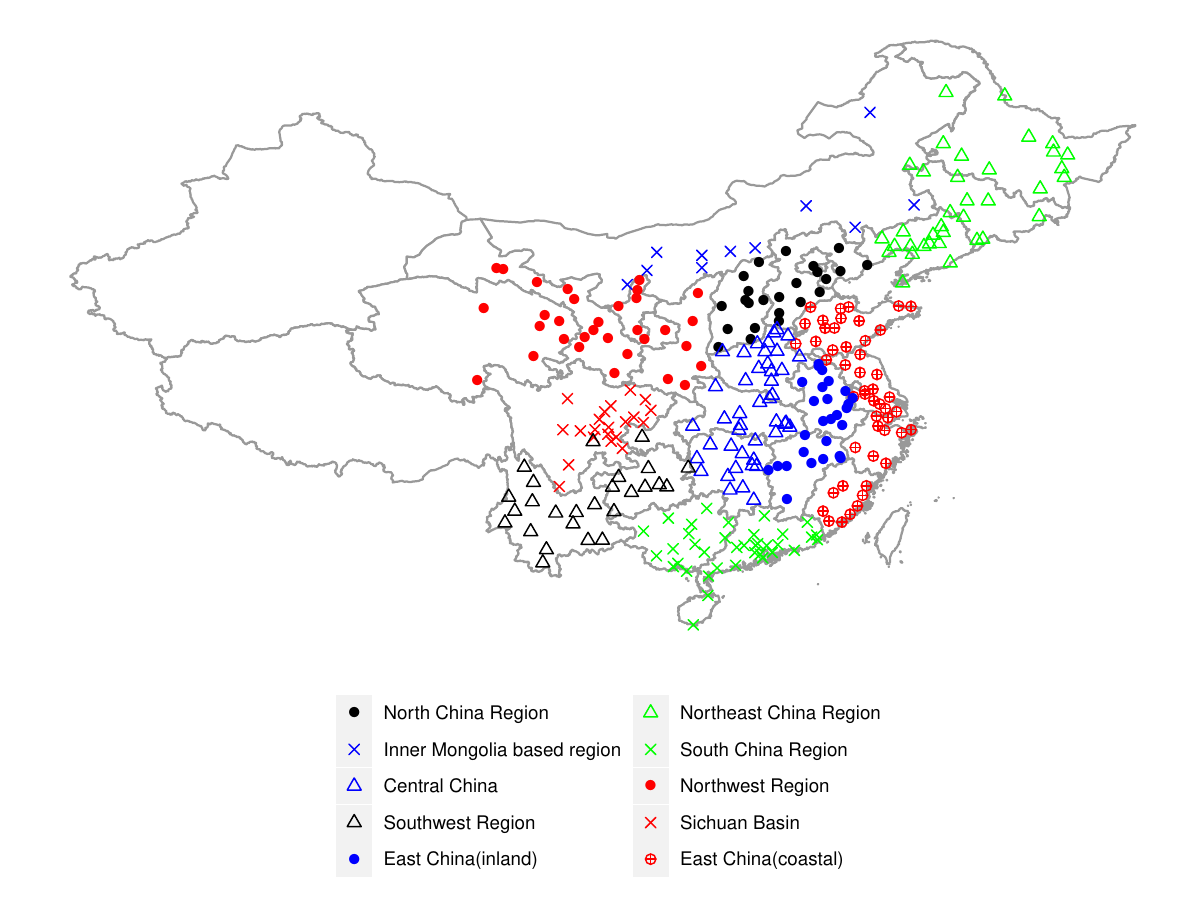}
      \caption{The input clusters. Input regionalization map of China's $\text{PM}_{2.5}$ of $307$ cities from January 2015 to December 2016. Each colored symbol represents a different cluster, and the total number of clusters is $10$.}
      \label{input}
      \end{center}
   \end{figure}

Now, we perform the clustering analysis of the $\text{PM}_{2.5}$ concentration data. In the initialization step, we generate the clustering results following the administrative divisions of China, which give the Northern China Region, Northeast China Plain, Northwestern Region, Southwestern Region, Southern China Region, Central China Region, and Eastern China Region. Three further breakdowns are made beyond these divisions. The first breakdown separates the Inner Mongolia Autonomous Region from the given seven main regions. This is because the Inner Mongolia Autonomous Region is the only province in China that belongs to multiple administrative regions. Treating Inner Mongolia as a separate cluster instead of dividing it into three parts is more convenient for policy-making, as provinces are smaller administrative units compared to administrative regions. The second breakdown divides the Eastern China region into two areas: the coastal area and the inland area. The reason for this division is that the distance of a city from the sea can impact its $\text{PM}_{2.5}$ concentration levels. Coastal cities tend to have lower $\text{PM}_{2.5}$ levels due to the influence of sea breezes and other factors. The third breakdown divides the Southwestern China region into the Sichuan Basin and other parts. An area with basin terrain usually has higher $\text{PM}_{2.5}$ levels because it is geologically surrounded by mountains or hills, which create a natural enclosure that may lead to the accumulation of pollutants in the air. The input clustering results indicate our belief that there are $10$ clusters with distinct $\text{PM}_{2.5}$ patterns. This aligns with the suggestion of environmental scientists \citep{wang2015spatial}, and we display the input clusters in Figure \ref{input}. 

 In the iteration step, we define the threshold of the distance between a given sample and its potential cluster to be the 2-percentile of between-sample distance for the distance restriction. We demonstrate in the appendices why the threshold is a good choice. It is worth pointing out that this restriction helps prevent the combination of geographically far apart sites since such a combination is practically meaningless.

\subsection{Clustering Analysis}\label{casestudy}

\begin{figure}[!ht]
      \begin{center}
      \includegraphics[width=1.0\textwidth]{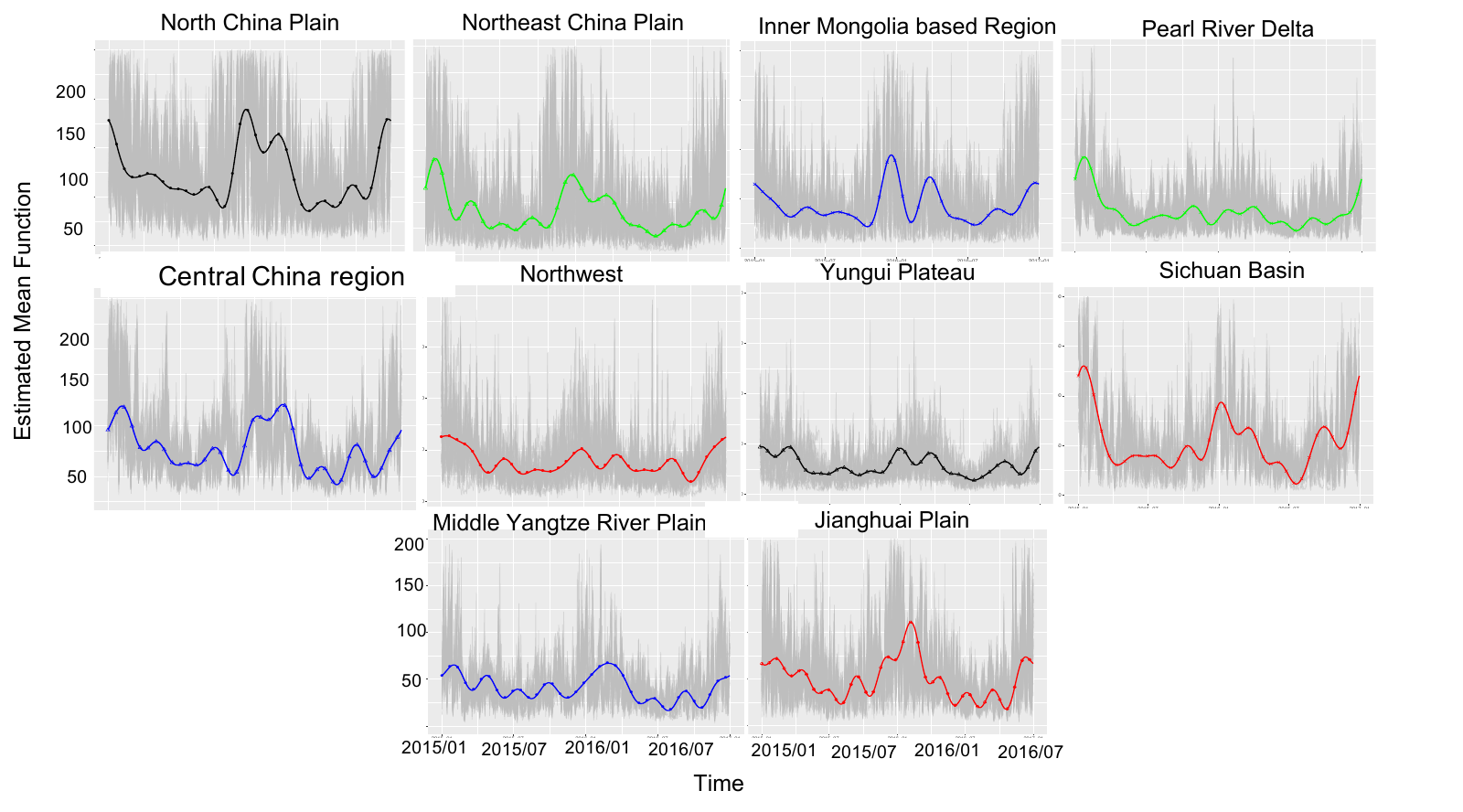} 
      \caption{Estimated mean function of ten clusters. The trimmed mean function of $\text{PM}_{2.5}$ levels of each output cluster. Same legend as in Figure \ref{output}, with the observed $\text{PM}_{2.5}$ concentrations are marked in grey.}
      \label{meanf}
      \end{center}
   \end{figure}

\begin{figure}[!ht]
      \begin{center}
      \includegraphics[width=0.6\textwidth]{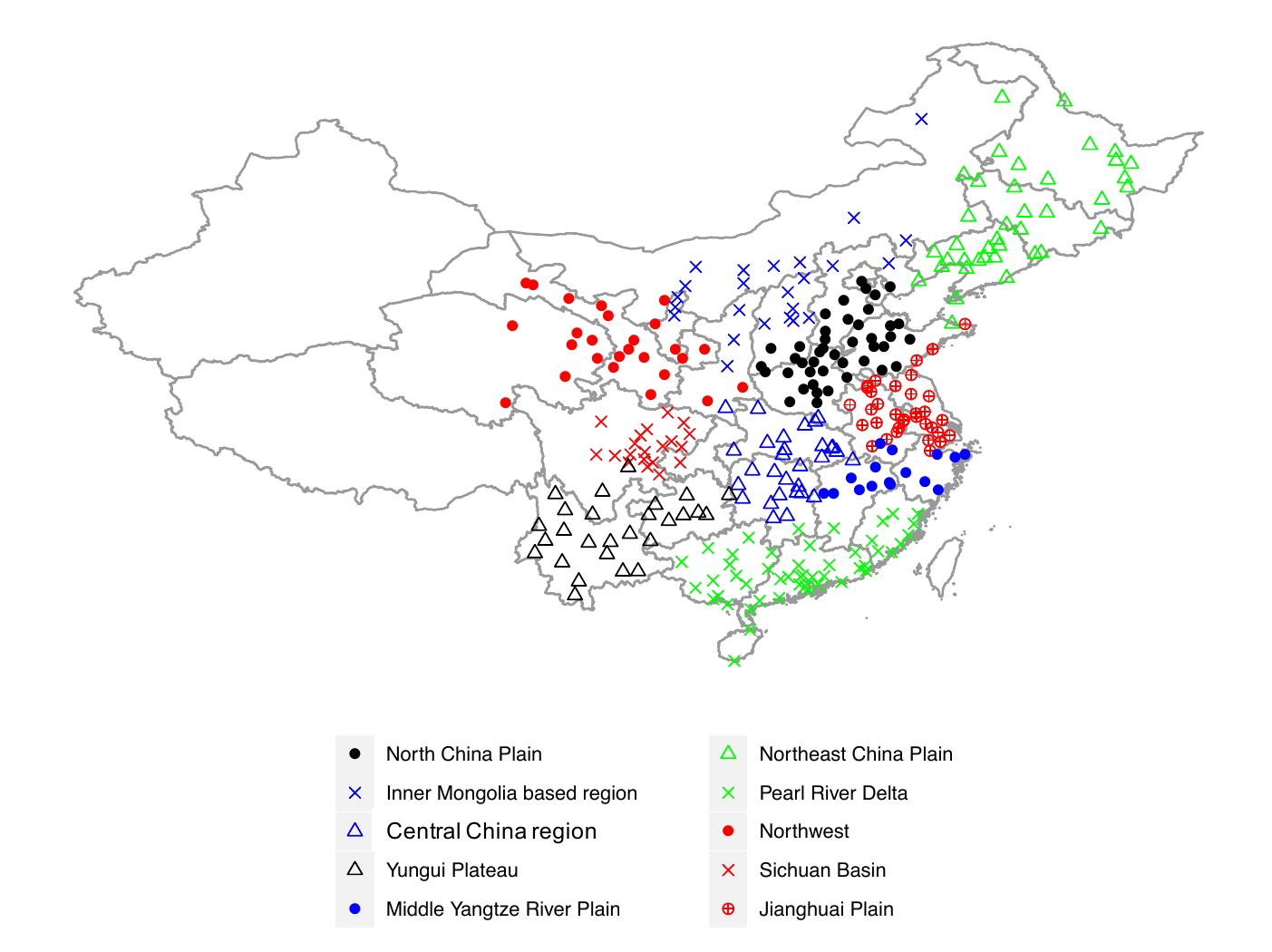} 
      \caption{The output clusters. Output regionalization map of China's $\text{PM}_{2.5}$ of $307$ cities from January 2015 to December 2016. Each colored symbol represents a different cluster, and the total number of clusters is $10$.}
      \label{output}
      \end{center}
   \end{figure}

We apply the RSTC clustering method to the city-based spatio-temporal $\text{PM}_{2.5}$ data and obtain ten clusters with clear boundaries. For each cluster, we use a cubic B-spline with 30 equally-spaced interior knots to model the trimmed mean functions, and the resulting curves are presented in Figure \ref{meanf}. We display the clustering results on the map of China in Figure \ref{output}.

The main seasonal patterns and spatial trends across the country are revealed in both figures. To be specific, the trimmed mean functions displayed in Figure \ref{meanf} show a consistent `W' shape despite the varying temporal trends across regions, which implies that the levels of $\text{PM}_{2.5}$ are typically higher during winter compared to a summer \citep{liang2021modeling}.  This pattern is more obvious in Northern and Central China but less noticeable in the southern regions. To be specific, a clear `W' shape is observed in the North China Plain, Northeast China Plain, Central China region, and Sichuan Basin. In contrast, the $\text{PM}_{2.5}$ levels remain consistently low in the Pearl River Delta and Yungui Plateau. The increase in $\text{PM}_{2.5}$ levels in winter is commonly attributed to coal burning and straw burning. In addition to seasonal patterns, a notable geographical trend is observed, with consistently higher levels of $\text{PM}_{2.5}$ concentration in Northern China compared to Southern China. This disparity can be attributed to various factors, including geographical factors, meteorological conditions, and local emissions, such as traffic and industrial emissions. However, some exceptions contradict the geographical trend. For example, the Northwestern China Region has consistently low $\text{PM}_{2.5}$ levels throughout the year. In contrast, the Sichuan Basin, which lies in Southern China, has surprisingly high $\text{PM}_{2.5}$ levels. Locally distinct features together may contribute to the situation. The estimated trimmed mean functions also reveal the decadal inconsistencies in local $\text{PM}_{2.5}$ trends. Specifically, the $\text{PM}_{2.5}$ levels in the Northern China Region and Inner Mongolia-based clusters during the winter of 2015 were much lower than in the winter of 2016, while the Southern China region, Southwestern China region, and Eastern China region exhibit an opposite trend. 

\begin{figure}[!ht]
      \begin{center}
      \includegraphics[width=1\textwidth]{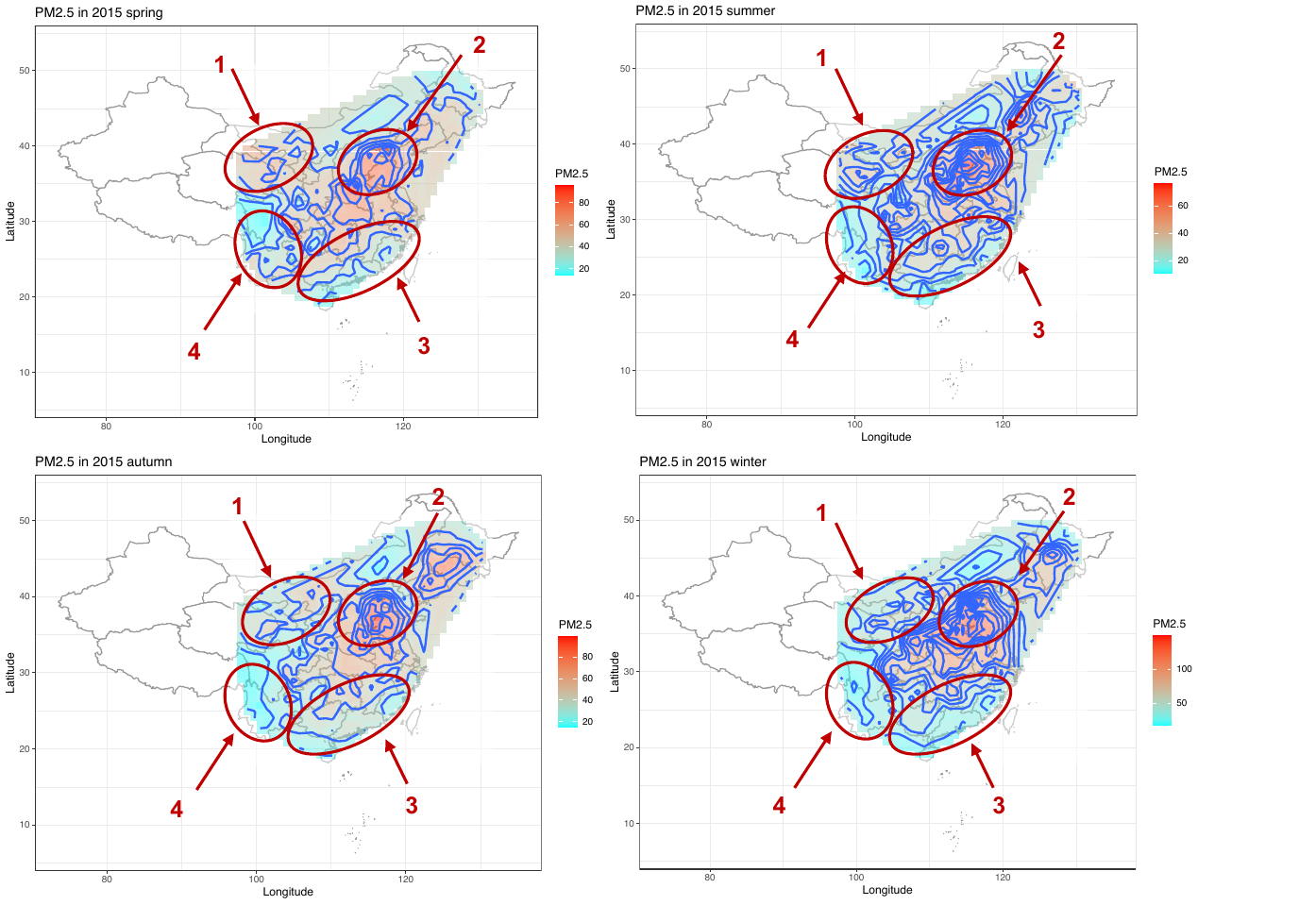} 
      \caption{The contour plot of mean daily $\text{PM}_{2.5}$ levels of each season in 2015 in China. The average daily $\text{PM}_{2.5}$ concentrations during the spring season were calculated based on data from March 2015 to May 2015; the ones during summer were calculated based on data from June 2015 to August 2015; the ones during autumn were calculated based on data from September 2015 to November 2015; and the ones during winter were calculated based on data from December 2015 to February 2016. The cluster labeled as "1" stands for the Northwest China, the cluster labeled as "2" stands for the North China Plain, the cluster labeled as "3" stands for the Pearl River Delta, and the cluster labeled as "4" stands for Yungui Plateau.} 
      \label{contour}
      \end{center}
   \end{figure}

Through the RSTC process, we are able to identify clusters with homogeneous $\text{PM}_{2.5}$ patterns. For better illustration, we display the contour plots of average daily $\text{PM}_{2.5}$ levels for each season of 2015 in Figure \ref{contour}. We find that the contours in the Pearl River Delta, the Yungui Plateau, and Northwest China are relatively open in all four seasons, indicating the uniformity of $\text{PM}_{2.5}$ levels within these regions throughout the entire year and further confirming the homogeneity within each of these identified clusters. In contrast, the North China Plain (NCP) exhibits distinct pollution centers characterized by closely packed contour lines, indicating the fast decline in $\text{PM}_{2.5}$ levels from the center to its surrounding areas. Despite the substantial variability, it is still reasonable to view it as a unified cluster. The existence of a pollution center within NCP and the decreasing $\text{PM}_{2.5}$ patterns in its vicinity indicate the existence of regional interaction effects. 

\section{DISCUSSION}\label{discussion}
{In this study, we establish a robust functional clustering framework. It is widely applicable to all sorts of situations and works effectively without much prior information relating to the data.} Also, we are able to illustrate the robustness of our methods through a noisy example, which shows our method can perform steadily under the existence of outliers when most methods fail to do so. We incorporate spatial considerations into the framework and apply it to a complex-structured and noisy dataset of $\text{PM}_{2.5}$ concentration around China. The clustering results generated by our method have clear and meaningful boundaries, which lay the foundation for policymaking. 

 To the best of our knowledge, this framework is the first one intended for spatial-temporal clustering and exhibiting robustness. We innovatively adopt angular depth in our framework, as the depth notion has proved to be a reliable tool in robust statistical analysis, and angular depth ensures the detection of magnitude and shape anomalies. Additionally, our framework is highly flexible as it can accommodate different regionalization inputs, and does not rely on specific model assumptions. It is also interpretable as the iteration process is a simple adaptive version of the K-means algorithm.
 
Through the established framework, we identify ten clusters with clear boundaries. $\text{PM}_{2.5}$ patterns exhibit homogeneity within each cluster and heterogeneity across clusters. The clustering results help reveal the underlying geographical and seasonal tendencies of $\text{PM}_{2.5}$ across the nation. The alignment between our results and the findings of previous studies validates the efficacy of our approach. For example, \citet{liang2021modeling} took the Inner Mongolia Region and the Northwestern region as a single entity, and we separated these two regions. This separation is supported by the evidence given in Figure \ref{meanf}, where discernible variations in pollution levels and distinct seasonal patterns emerge between the Inner Mongolia cluster and the Northwest China Region. This separation enables the policymakers to develop policies that are suitable to local regions and hence more effective in both environmental protection and economic development. This  trade-off  is a very important policy problem \citep{he2018study}.

Despite the aforementioned contributions, our research also sheds light on some new research questions. For instance, one important question is how to incorporate local characteristics such as meteorological factors, geographical features, and emissions patterns into the current process and propose an integrated clustering system. Although the patterns of particulate matter are the most important reference for clustering, it is still of scientific importance to have a systematic method to combine them with multiple local features. Another question is how to evaluate the uncertainty of cluster assignments. The measure of uncertainty is a statistical inference process, which is currently not involved in our process but can be of practical value. Furthermore, it is important to acknowledge that the levels of $\text{PM}_{2.5}$ can significantly fluctuate over the span of decades. Currently, our clustering results are generated using the $\text{PM}_{2.5}$ levels from two specific years. In future analyses, we should incorporate a more extensive range of $\text{PM}_{2.5}$ data and examine the chronological changes in clustering outcomes. This will allow us to draw conclusions regarding the decadal variance and adjust air pollution regulation policies accordingly.

In all, we believe the proposed robust spatial-temporal framework is of great value. It provides clear and meaningful regionalization results based on the $\text{PM}_{2.5}$ levels across the country. The results serve as important guidelines for the establishment of regional collaborative management of air pollution. The method itself provides a useful tool for regionalization based on a time series across a specific region. The method has the advantage of requiring no model assumptions, which brings enlightenment to the new area of robust clustering for spatial functional data. Still, the aforementioned questions relating to the framework are in need of further investigation.

%\section{BibTeX}
\normalem
\bibliographystyle{chicago}
\bibliography{Bibliography-MM-MC}
\end{document}